\documentclass[a4size]{article}
\usepackage[english]{babel}
\usepackage[utf8x]{inputenc}
\usepackage[T2A,T1]{fontenc}
\usepackage[a4paper,top=3cm,bottom=2cm,left=3cm,right=3cm,marginparwidth=1.75cm]{geometry}

\usepackage{amsmath}
\usepackage{listings}
\usepackage{graphicx}
\usepackage{cite}
\usepackage[colorinlistoftodos]{todonotes}
\usepackage[colorlinks=true, allcolors=blue]{hyperref}
\usepackage{authblk}
\title{\bf{Entropy of hard square lattice gas with $k$ distinct species of particles: coloring problems and vertex models}}
\author[1]{Sahil K. Singh}
\author[2,3,4]{Sudhir R. Jain}
\affil[1]{\it{Department of Physics, Indian Institute of Technology, Kanpur 208016, India}}
\affil[2]{\it{Nuclear Physics Division, Bhabha Atomic Research Centre, Mumbai 400085, India}}
\affil[3]{\it{Homi Bhabha National Institute, Training School Complex, Anushakti Nagar, Mumbai 400094, India}}
\affil[4]{\it{UM-DAE Centre for Excellence in Basic Sciences, Vidyanagari Campus, Kalina, Mumbai 400098, India}}

\begin{document}
\maketitle
\begin{abstract}
Coloring the faces of a two-dimensional square lattice with $k$ distinct colors such that no two adjacent faces have the same color is considered by establishing connection between the $k$-coloring problem and a generalized vertex model. Associating the colors with $k$ distinct species of particles with an infinite repulsive force between nearest neighbors of the same type and zero chemical potential $\mu$ associated with each species, the number of ways $[W(k)]^N$ for large $N$ is related to the entropy of the {\it{hard square lattice gas}} at close packing of the lattice, where $N$ is the number of lattice sites. We discuss the evaluation of $W(k)$ using transfer matrix method with non-periodic boundary conditions imposed along at least one direction, and, show the characteristic Toeplitz block structure of the transfer matrix. Using this result, we present some analytical calculations for non-periodic models that remain finite in one dimension. The case $k = 3$ is found to approach the exact result obtained by Lieb for the residual entropy of ice with periodic boundary conditions. Finally, we show, by explicit calculation of the contribution of subgraphs and the series expansion of $W(k)$, that the generalized Pauling type estimate (which is based on mean-field approximation) dominates at large values of $k$. We thus also provide an alternative series expansion for the chromatic polynomial of a regular square graph.
\end{abstract}

\noindent
Keywords: vertex model; coloring problem; transfer matrix; entropy

\section{Introduction}

Ice has a residual entropy at low temperatures that cannot be explained by any conceivable lattice vibrations \cite{pauling1,pauling2,lieb}. Pauling \cite{pauling1,pauling2,bernal,lieb} suggested that this is because the O-O bond length (2.76 \AA) is more than twice the O-H bond length (0.95 \AA) and thus the hydrogen atom has two possible positions on each of the O-O bond. ``In the gas molecule the O-H distance is 0.95 \AA , and the  magnitudes of the changes in properties from steam to ice are not sufficiently great to permit us to assume that this distance is increased to 1.38 \AA ", writes Pauling \cite{pauling1}. Furthermore, he writes ``The concentration of $(\text{OH})^-$ and $(\text{H}_3\text{O})^+$ ions in water is very small and we expect the situation to be unchanged in ice. Thus each oxygen atom must be surrounded by two oxygen atoms near it and two on the far side" \cite{pauling1,bernal,lieb}.  Associating with each hydrogen atom near (far from) an oxygen atom an arrow pointing towards (away from) the oxygen atom , the problem can be transformed into an interesting graph-theoretic problem of determining the number of ways of arranging arrows on the edges of square lattice such that the number of arrows going into a vertex is equal to that coming out of the vertex \cite{thompson}. Lenard(according to \cite{lieb}) has shown that, aside from a factor of three, the problem is isomorphic to the three colorings of the square lattice. The problem of residual entropy of ice and related problems has been discussed extensively in the literature\cite{pauling1,pauling2,baxter1,baxter2,runnels,lieb,thompson,stillinger,nagle}. Lieb \cite{lieb} found the exact value for the exponent $W(3)$ for a toroidal lattice as:
\begin{equation*}
 W(3)={\left(\frac{4}{3}\right)}^{3/2}\approx 1.5396\ldots
\end{equation*}
Baxter \cite{baxter1,baxter2} found the grand canonical partition function for the 3-coloring of the square lattice where the activities $z_1,z_2,z_3$ are associated with each color and every configuration at a vertex has zero energy, which is equivalent to a general case of the close packed hard square lattice gas mentioned before. Thus Lieb's result \cite{lieb} is recovered when $z_1=z_2=z_3=1$. Pauling \cite{pauling2} made an approximation for $W$ based on the assumption that the lattice bonds and vertex configurations are independent. If there are $N$ lattice sites, then the total number of ways of placing arrows,  were there no restrictions,  would be $2^{2N}$, because there are $2N$ bonds in the lattice. Because of the ice condition, only 6 out of 16 possible configurations are allowed at each vertex. Thus the number of ways is:
\begin{equation*}
W_N \approx 2^{2N}\left(\frac{6}{16}\right)^N
\end{equation*}
and thus \begin{equation*}
W=\lim_{N\rightarrow\infty}{W_N}^{1/N}\approx\frac{3}{2}=1.5
\end{equation*}
The above estimate is due to Pauling. Nagle \cite{nagle} improved the methods of Stillinger et al. \cite{stillinger} to obtain the series expansion of $W$:\begin{equation*}
W=\frac{3}{2}\left(1+\frac{1}{3^4}+\frac{4}{3^6}+\frac{22-4}{3^8}+\ldots \right)
\end{equation*} 
In this paper, we construct a vertex model which maps to the $k$ coloring of the square lattice. We begin by discussing transfer matrix method which incorporates the interaction of two rows of a lattice. The transfer matrix for chains with {\it non-periodic} boundary condition imposed on the rows is found to have Toeplitz block structure. We also show that the maximum eigenvalue of the transfer matrix for non-periodic boundary conditions approaches the result obtained by \cite{lieb} for $k=3$ for toroidal ice, as expected. Analytical values of open and cylindrical square lattices that remain finite in only one dimension is derived (for the case $k=3$) without resorting to the method of diagonalizing the transfer matrix. We illustrate the $k$-coloring problem by considering in detail the case of $k = 4$ for brevity where we find a 21-vertex model. Graph theoretical methods \cite{stillinger} and matrix methods \cite{nagle} are generalized to obtain a series expansion of $W(k)$ in terms of $k$. The first term of the series is the generalized Pauling estimate and subsequent terms are $\mathcal{O}({1/{k^n}})$. The results of \cite{nagle} is recovered for the special case of $k=3$, corresponding to the six-vertex model.

\section{The transfer matrix approach}
Consider a square lattice with $m$ rows, each having $p$ lattice sites. The last lattice site of each row is connected to the first lattice site via a bond and the $m^{th}$ row is connected to the first row via $p$ bonds. Thus, while considering the assignment of arrows on the bonds of the lattice, we have $m$ rows of $p$ up-down arrows and $p$ left-right arrows. Let $\phi$ denote a possible configuration of a row of $p$ vertical bonds. For two adjacent rows of vertical bonds having configurations $\phi$ and $\phi'$, define $B(\phi,\phi')$ to be the number of ways of arranging horizontal arrows on the row of horizontal bonds common to these two rows. Thus $B$ is a $2^p \times 2^p$ matrix called the transfer matrix \cite{baxter2} and the number of ways $Z$ of arranging the arrows on the lattice is: 

\begin{equation*}
Z=\sum_{\phi_{1}}\ldots \sum_{\phi_{m}}B(\phi_1,\phi_2)B(\phi_2,\phi_3)\ldots B(\phi_{m-1},\phi_m)B(\phi_m,\phi_1) = Tr[B^m].
\end{equation*}  
For large $m$, 
\begin{equation*}
Z=\Lambda^m.
\end{equation*}
where $\Lambda$ is the maximum eigenvalue of $B$.
\subsection{The {\it{non-toroidal}} ice}
Unlike Lieb\cite{lieb}, we consider a square lattice that is not toroidal, i.e., either it is a completely open square lattice or a cylindrical lattice with the last row connected to the first row.
\subsubsection{Cylindrical lattice}\label{cylindricalice}
For a cylindrical lattice, we have 
\begin{equation*}
Z=\sum_{\phi_{1}}\ldots \sum_{\phi_{m}}B(\phi_1,\phi_2)B(\phi_2,\phi_3)\ldots B(\phi_{m-1},\phi_m)B(\phi_m,\phi_1) = Tr[B^m].
\end{equation*}
Note that the matrix $B$ is not the same as that for the toroidal ice because removing the cyclic boundary condition in each row would allow some of the configurations of two adjacent rows to have non-zero elements $B(\phi,\phi')$ which were zero for the cyclic rows.

Let $\mathcal{L}_p$ be an ordered set of all possible configurations of $p$ up-down arrows. Thus the size of $\mathcal{L}_p$ is $2^p$. Let
\begin{equation}
\mathcal{L}_1=\left\lbrace\uparrow,\downarrow\right\rbrace.
\end{equation} 
The set $\mathcal{L}_p$ is constructed in the following manner: \begin{equation}
\text{If } \mathcal{L}_p=\left\lbrace\phi_1,\phi_2,\ldots,\phi_{2^p}\right\rbrace\text{, then } \mathcal{L}_{p+1}=\left\lbrace\uparrow\phi_1,\uparrow\phi_2,\ldots,\uparrow\phi_{2^p},\downarrow\phi_1,\downarrow\phi_2,\ldots,\downarrow\phi_{2^p}\right\rbrace. 
\end{equation}
where $\uparrow\phi$ denotes a configuration of $p+1$ arrows whose first arrow is up and the rest are in configuration $\phi$, and similarly for $\downarrow\phi$. Note that the first $2^p$ elements of $L_{p+1}$ have their first arrow as $\uparrow$ and the next $2^p$ have their first arrow as $\downarrow$, and this observation is very crucial in the arguments that follow. Let $B_p(i,j)=B(\phi_i,\phi_j)$ where $\phi_i,\phi_j\in\mathcal{L}_p$ and $\phi_i$ is the $i^{th}$ element of $\mathcal{L}_p$. Also, 
\begin{equation}\label{dcmp1}
B(\phi,\phi')=B(\downarrow\phi,\uparrow\phi')+B(\uparrow\phi,\downarrow\phi').
\end{equation}

\begin{figure}
\centering
\includegraphics[width=0.5\textwidth]{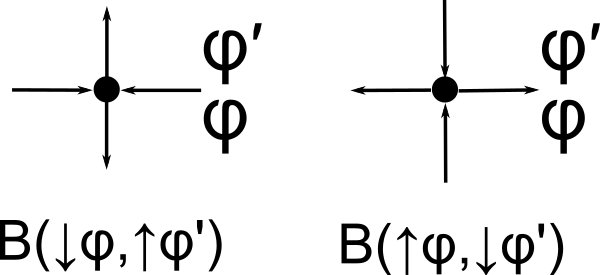}
\label{fig:arrow}
\caption{Configurations counted by $B(\downarrow\phi,\uparrow\phi')$ and $B(\uparrow\phi,\downarrow\phi')$}
\end{figure}
The proof of Equation \ref{dcmp1} is simple: the downarrow-uparrow in the first sites of two adjacent rows would force the horizontal arrow at the beginning of $\frac{\phi'}{\phi}$ to take the left position($\leftarrow$), which is counted by the first term of right hand side, the uparrow-downarrow in the first sites of two adjacent rows would force the horizontal arrow at the beginning of $\frac{\phi'}{\phi}$ to take the right position($\rightarrow$), which is counted by the second term(refer to Fig. 1). Since the horizontal arrow at the beginning of $\frac{\phi'}{\phi}$ can point either to the left or to the right, all possibilities are exhaustively counted by the right hand side of the equation.
Define\begin{equation}\label{dcmp}
B(\downarrow\phi,\uparrow\phi')=A(\phi,\phi').
\end{equation}
Let $A_p(i,j)=A(\phi_i,\phi_j)$ where $\phi_i,\phi_j\in\mathcal{L}_p$ and $\phi_i$ is the $i^{th}$ element of $\mathcal{L}_p$. Since $B(\phi,\phi')=B(\phi',\phi)$ \begin{equation}
B_p(i,j)=A_p(i,j)+A_p^T(i,j).
\end{equation}
 Now, we consider the four $2^p\times2^p$ block matrices of the $2^{p+1}\times2^{p+1}$ matrix $A_{p+1}$. For the upper left block ($1{\leq}i{\leq}2^p$ and $1{\leq}j{\leq2}^p$), we have: \begin{equation}
 A_{p+1}(i,j)=A(\uparrow\phi_i,\uparrow\phi_j)=B(\downarrow\uparrow\phi_i,\uparrow\uparrow\phi_j)=B(\downarrow\phi_i,\uparrow\phi_j)=A_p(i,j).
 \end{equation}
 where $\phi_i,\phi_j\in\mathcal{L}_p$. 
 For the lower right block($2^p<i\leq2^{p+1}$ and $2^p<j\leq2^{p+1}$), we have:
 \begin{equation}
 A_{p+1}(i,j)=A(\downarrow\phi_{i-2^p},\downarrow\phi_{j-2^p})=B(\downarrow\downarrow\phi_{i-2^p},\uparrow\downarrow\phi_{j-2^p})=B(\downarrow\phi_{i-2^p},\uparrow\phi_{j-2^p})=A_p(i-2^p,j-2^p)
 \end{equation}
where $\phi_{i-2^p},\phi_{j-2^p}\in\mathcal{L}_p$. Thus the diagonal blocks of $A_{p+1}$ are the same as $A_p$. For the upper off-diagonal block($1{\leq}i{\leq}2^p$ and $2^p<j\leq2^{p+1}$), we have: \begin{equation}
 A_{p+1}(i,j)=A(\uparrow\phi_{i},\downarrow\phi_{j-2^p})=B(\downarrow\uparrow\phi_i,\uparrow\downarrow\phi_{j-2^p})=B(\uparrow\phi_i,\downarrow\phi_{j-2^p})=A_p^T(i,j-2^p).
 \end{equation}
 where $\phi_i,\phi_{j-2^p}\in\mathcal{L}_p$.
 By similar argument, the lower off-diagonal block has all elements $0$. Thus
 \begin{equation}\label{eq:turner}
 A_{p+1}=\begin{pmatrix}
 A_p & A_p^T \\
 0 & A_p
 \end{pmatrix}
 \end{equation}
 We have the $1\times1$ matrix $A_0=B(\downarrow,\uparrow)=[1].$  Eq. \eqref{eq:turner} sets up a recursive scheme whereby as $p$ increases, we can find the maximum eigenvalue, and recover the results by Lieb. For the triangular lattice also, arguments as above lead to block matrices and whence ensue the corresponding results by Baxter. 
 
\subsection{Lieb's square ice constant for cylindrical chain}
From the matrices constructed in the previous section, it is clear that
\begin{equation*}
\lim_{p\rightarrow\infty}[\lambda_{max}(A_p+A_p^T)]^{1/p}
\end{equation*}
gives Lieb's result.
We performed the computation of $[\lambda_{max}(A_p+A_p^T)]^{1/p}$ till $p=10$ using MATHEMATICA and the values we found were the following: 
\begin{center}
 \begin{tabular}{||c c||} 
 \hline
 $p$ & $[\lambda_{max}(A_p+A_p^T)]^{1/p}$ \\ [0.5ex] 
 \hline\hline
 1 & 3.00000 \\ 
 \hline
 2 & 2.13578 \\
 \hline
 3 & 1.91037 \\
 \hline
 4 & 1.80789\\
 \hline
 5 & 1.74955 \\
 \hline
 6 & 1.71195 \\
 \hline
 7 &  1.68573 \\
 \hline
 8 &  1.66641 \\
 \hline
 9&  1.65159 \\
 \hline
 10 & 1.63987 \\ [1ex] 
 \hline
\end{tabular}
\end{center}
A plot of these data points is shown in Fig. 2. We fitted the data to the following functional form:
\begin{equation*}
a_0+a_1/x+a_2/{x^2}+\ldots+a_5/{x^5}
\end{equation*}
We obtained the asymptotic limit as $a_0=1.53967$, which is not too bad all things considered.

\begin{figure}
\centering
\includegraphics[width=0.5\textwidth]{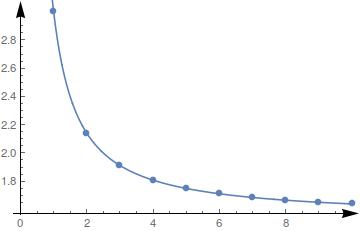}
\caption{Plot of $[\lambda_{max}(A_p+A_p^T)]^{1/p}$ vs. $p$ for $p=1,2,3....,10$}
\end{figure}

 We prove two theorems involving the matrix $A_p$ that will be useful later.
 \newtheorem{theorem}{Theorem}
 \begin{theorem}
 $Tr[A_p^n]=2^p$ for all non-zero positive integers $n$ and $p$.
 \end{theorem}
 It is elementary to show that
 \begin{equation}
 A_{p+1}^n=\begin{pmatrix}
 A_p^n & X_n \\
 0 & A_p^n
 \end{pmatrix}
 \end{equation}
 where $X_n=\sum_{i=1}^nA_p^{n-i}A_p^TA_p^{i-1}$, with $A_p^0$ taken to be equal to $I_p$.
 Thus 
 \begin{equation}
 Tr[A_{p+1}^n]=2Tr[A_{p}^n].
 \end{equation}
 Since $Tr[A_0^n]=1$, we get 
 \begin{equation}
 Tr[A_p^n]=2^p.
 \end{equation}
 
 \begin{theorem}
 $Tr[A_p^nA_p^T]=(n+2)^p$ for all non-zero positive integers $n$ and $p$.
 \end{theorem}
 
 \begin{equation}
 A_{p+1}^T=\begin{pmatrix}
 A_p^T & 0\\
 A_p & A_p^T
 \end{pmatrix}.
 \end{equation}
 and thus
 \begin{equation}
 A_{p+1}^nA_{p+1}^T=\begin{pmatrix}
 A_p^nA_p^T+X_nA_p & X_nA_p^T\\
 A_p^{n+1} & A_p^nA_p^T
 \end{pmatrix}.
 \end{equation}
 It is trivial to show, by using cyclic property of trace of product of matrices, that
 \begin{equation}
 Tr[X_nA_p]=nTr[A_p^nA_p^T].
 \end{equation}
 We get the following iterative formula
 \begin{equation}
 Tr[A_{p+1}^nA_{p+1}^T]=(n+2)Tr[A_p^nA_p^T]
 \end{equation}
 since $Tr[A_0^nA_0^T]$=1, we have
 \begin{equation}
 Tr[A_p^nA_p^T]=(n+2)^p.
 \end{equation}
 Equipped with the block matrix structure and the above two theorems, we present the exact number of ways of satisfying the ice condition of a $2\times p$ and $3\times p$ cylindrical lattice in the Appendix, for illustrative purpose.  

\subsection{The 4-coloring problem : ``warm up"}
The connection between the ice problem and the 3-coloring problem was argued by Andrew Lenard (it appears in \cite{lieb}). We present the case of four colors, which we hope will serve as an illustration for the general case discussed later.

\begin{figure}
\centering
\includegraphics[width=0.4\textwidth]{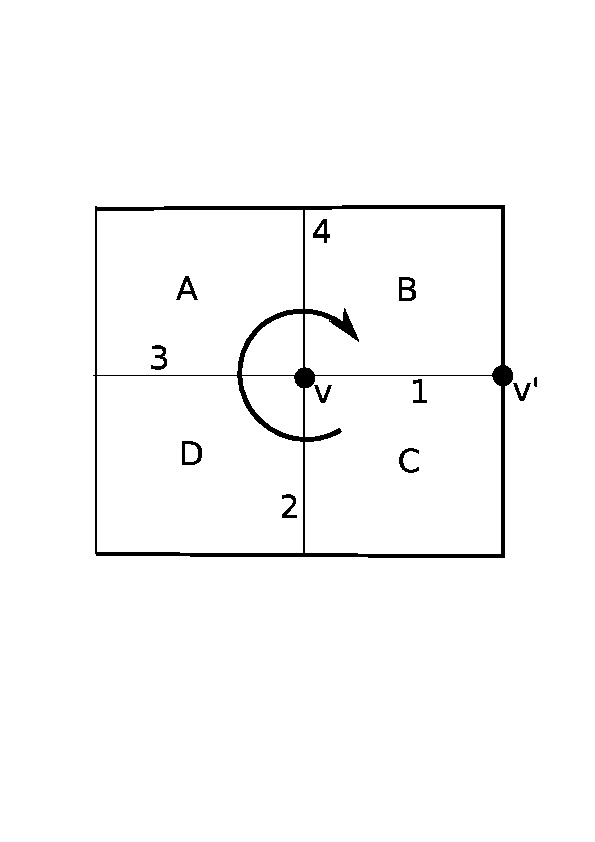}
\caption{About a vertex v, we have four faces colored by A, B, C, D, each of these being one of the four colors (coded using numbers $\in \{ 1, 2, 3, 4 \}$). To assign a state to an edge 1 (for instance), we traverse about the vertex v in a clockwise direction and notice that C comes after B. Edge 1 is then in a state $(C - B) \mod 4$. With respect to the vertex v' though, the state of edge 1 will be $(B - C) \mod 4$, because while going clockwise around v', the color B comes after color C. This illustrates the two rules, \eqref{arrow}, \eqref{icecondition} for the case of 4-coloring. The marks 1, 2, 3, 4 in the figure denote the edges, not to be confused with colors or states of edges.}
\end{figure}

Let the four colors be given numbers: ${\mathbf S} = \{1, 2, 3, 4 \}$. The edges emanating from a given vertex can be assigned states by traversing in (say) clockwise direction about it, in terms of the two distinct colours on two sides of an edge. Precisely, $A, B, C, D$ are colors (they could be any $\in {\mathbf S}$ shown in the Fig. 3, with $D \neq A \neq B, B \neq C \neq D$. On traversing clockwise about the vertex $v$, we assign the number $(C - B) \mod 4$  to edge 1. Clearly, the state of edge 1 with respect to the vertex $v'$ is $(B - C) \mod 4$ in such a way that $(C - B) \mod 4 + (B - C) \mod 4 \equiv 0 \mod 4$. To abbreviate the notation, the states of edge 1 w.r.t. vertex $v$ ($v'$) are $e_1$ ($e_1'$). Let the states of edges 2,3 and 4 w.r.t. $v$ be $e_2$, $e_3$ and $e_4$ respectively. It is obvious now that, $e_1 + e_2 + e_3 + e_4 \equiv 0 \mod 4$ and $e_1+e_1'\equiv 0 \mod4$. These rules will hold at every vertex of the lattice. The edge states are uniquely specified for a given coloring, because the $\mod 4$ operation will always give us a unique number.(The positive remainder we get by dividing an integer by 4 is always unique). Corresponding to every assignment of states satisfying the two rules above, there are four ways of coloring the faces of the square lattice using four colors. Let us start with a face and color it by, say, $A$. We could have chosen any one of the four colors, and it is this freedom that gives us a factor of four. If we traverse about $v$ clockwise(where $v$ is one of the vertices of face colored $A$) and use the color $X$($X \neq A$) for the next face, sharing the edge, then $X$ must satisfy $(X - A) \mod 4 \equiv e_1$(where $e_1$ is the state of common edge w.r.t. $v$). That $X$ is unique can be immediately proved by assuming the contradiction that there are two colors, $X_1, X_2$ obeying 
$(X_1 - A) \mod 4 \equiv e_1$ and $(X_2 - A) \mod 4 \equiv e_1$. Immediately, we have $X_1 = X_2$ as $(X_1 - X_2) \mod 4 \equiv 0$ and $X$'s can only take values from the set, ${\mathbf S}-\left\lbrace A \right\rbrace$. This also shows that if we take any two different $X$s and calculate the value $(X-A)\mod 4$ for these two $X$s, we will never get the same value. Thus out of all the three possibilities of $X$, one (and only one) will exist that will satisfy $(X-A)\mod 4=e_1$  In this way, we can color all the faces for a given edge state configuration. 

Let us denote the number of ways of assigning edge states to a square lattice having $N$ lattice sites by $W_{\rm vertex}$, and, the number of ways of coloring the faces by $W_{\rm coloring}$. We see that $W_{\rm coloring} = 4W_{\rm vertex}$; however, the factor `4' is immaterial when we consider the quantity $\lim_{N \to \infty } W^{1/N}$. 

The number of possible configurations such that $e_1 + e_2 + e_3 + e_4 \equiv 0 \mod 4$ is $(4-1)^3 - (4-1)^2 + (4-1) = 21$. Thus, we have established that 4-coloring of a square lattice is equivalent to a 21-vertex model (Fig. 4). 

\begin{figure}
\centering
\includegraphics[width=1.0\textwidth]{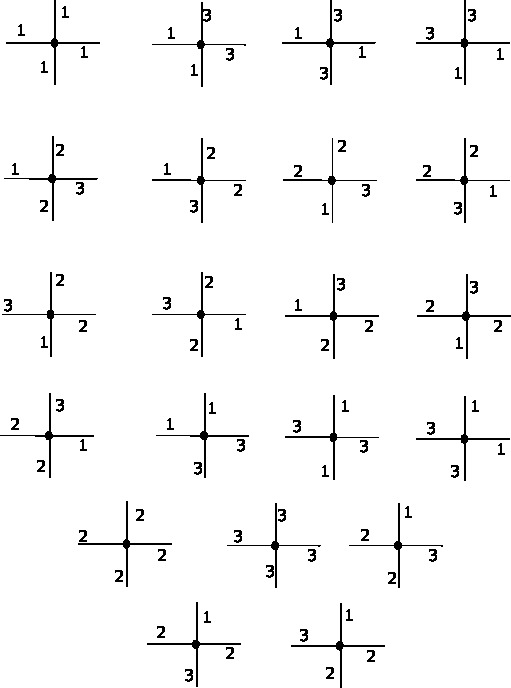}
\caption{21 vertex configurations corresponding to 4-coloring problem}
\end{figure}

\begin{figure}
\centering
\includegraphics[width=0.8\textwidth]{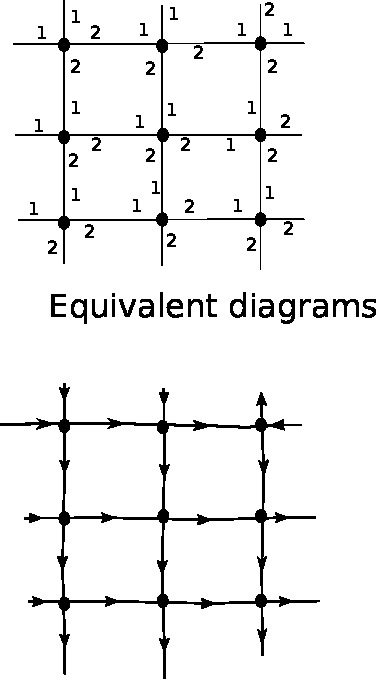}
\caption{An illustrative instance showing equivalence between 3-coloring on a square lattice and six-vertex model. Here we have adopted the convention that if the state of an edge with respect to a vertex is 1(2), then we draw an arrow pointing towards(away from) that vertex.}
\end{figure}

\subsection{The $k$-coloring problem}\label{color}
  
 The generalization of the above scheme to $k$-coloring immediately follows. We number the colors as $\left\lbrace1,2,\ldots,k\right\rbrace$. We follow the same procedure for assigning edges states as for the 4 coloring problem but we replace the $\mod 4$ with $\mod k$ (For example edge 1 in Fig. 3 is assigned the state $(C-B)\mod k$ w.r.t. vertex $v$). By construction, the states of the edges will belong to the set $\mathcal{F}_k-\left\lbrace0\right\rbrace$, where $\mathcal{F}_k$ is a set such that
 \begin{equation*}
  \mathcal{F}_k=\left\lbrace0,1,2,\ldots,{k-1}\right\rbrace.
 \end{equation*}
 The states will then satisfy the following conditions:\\
 \begin{enumerate}
 
 \item If $v_1$ and $v_2$ are two vertices connected by an edge $e$, and $e_1$ and $e_2$ are the respective states of $e$ with respect to $v_1$ and $v_2$, then 
 \begin{equation}\label{arrow}
 e_1+e_2=0\mod{k}
 \end{equation} where $e_1,e_2\in\mathcal{F}_k-\left\lbrace0\right\rbrace$.

 \item If $e_1,e_2,e_3$ and $e_4$ are the states (with respect to vertex $v$) of four edges emanating from the vertex $v$ (Fig. 3), then\begin{equation}\label{icecondition}
 e_1+e_2+e_3+e_4=0\mod{k}
 \end{equation}
 where $e_1,e_2,e_3,e_4\in\mathcal{F}_k-\left\lbrace0\right\rbrace$.
 \end{enumerate}
It is very easy to show that corresponding to every edge state configuration satisfying the above two conditions, there exists "$k$ possible $k$ coloring" of the faces.(Like in the case of 4 coloring, we can always find a unique color $X$($\neq A$) satisfying $(X-A)\mod k=e_1$, where $e_1\in\mathcal{F}_k-\left\lbrace0\right\rbrace$.)
 In fact, for $k=3$, the problem stated above is just the ice problem in disguise. To show how this works for the 3-coloring and associated six-vertex model, we show an example for illustrative purpose (Fig. 5).
 \subsubsection{The transfer matrix for open rows}
 The transfer matrix for the problem can be written in a similar manner as for the classic ice problem by Lieb. Consider two adjacent rows with $p$ lattice sites each, labelled $r_1$ and $r_2$ ($r_2$ above $r_1$). Let $\phi$ be a configuration of $p$ states of the row of vertical edges just below $r_1$ with respect to the corresponding lattice sites of $r_1$, and $\phi'$ be the configuration of $p$ states of the row just below $r_2$ with respect to the corresponding lattice sites of $r_2$. $B(\phi,\phi')$ is the number of ways of assigning states on the row of horizontal edges common to these two rows such that the conditions embodied in  \eqref{arrow} and \eqref{icecondition} are satisfied. We argue that it is possible to choose and fix configurations $\Phi(n)$ and $\Phi'(n)$ such that the state of the horizontal edge hanging out of the last lattice site of $\frac{\Phi'(n)}{\Phi(n)}$ will take the value $n$ and no other value, $n\in\mathcal{F}_k-\left\lbrace0\right\rbrace$ (refer to Fig. 4). Let $(a_i+b_i)\mod{k}=A_i$, $A_i\neq0\mod{k}$ for all $i$. The state of the second horizontal edge with respect to the second lattice site can have all possible states except $A_1$. The state of the third horizontal edge with respect to the third lattice site can have all possible states except $A_1+A_2$ and $A_2$. The state of the $i^{th}$ horizontal edge with respect to the $i^{th}$ lattice site can have all possible values except $(A_1+A_2+\ldots+A_{i-1}),(A_2+A_3+\ldots+A_{i-1}),\ldots,A_{i-1}$(all numbers are $\mod{k}$). Thus by suitably choosing the numbers $A_i$, it is possible to exclude $i-1$ states from the $i^{th}$ horizontal edge. Similarly, it is possible to exclude $i$ distinct states from the ${i+1}^{th}$ horizontal edge. Thus if $i=(k-2)$ and we suitably choose the numbers $A_i$, then we can exclude all states except $(k-n)$. Thus the state of the last horizontal edge in the figure with respect to the last lattice site can only have the state $n$, and our argument is complete.
 \begin{figure}
 \centering
 \includegraphics[width=0.7\textwidth]{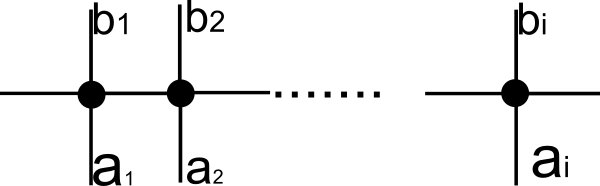}
 \caption{Our premise for choosing $\Phi(n)=\left\lbrace a_1,a_2,\ldots a_i\right\rbrace$ and $\Phi'(n)=\left\lbrace b_1,b_2,\ldots,b_i\right\rbrace$ mentioned in the text}
 \end{figure}
 Define 
 \begin{equation}
 A_n(\phi,\phi')=B(\Phi(n)\phi,\Phi'(n)\phi') 
 \end{equation}
 which is the generalization of \eqref{dcmp}. Define $A_n(\phi,\phi')=0$ when $n=0\mod{k}$, because the state of an edge cannot be $0$ with respect to any vertex.
 $\Phi(n)\phi$ denotes configuration $\phi$ appended to $\Phi(n)$. The state of the last horizontal edge of $\frac{\Phi'(n)}{\Phi(n)}$ with respect to the first lattice site of $\frac{\phi'}{\phi}$ in $\frac{\Phi'(n)\phi'}{\Phi(n)\phi}$ is $-n\mod{k}=k-n$. As $n$ varies from $1$ to $k-1$, $k-n$ will vary from $k-1$ to $1$ and the state of the first horizontal edge of $\frac{\phi'}{\phi}$ with respect to the first lattice site of $\frac{\phi'}{\phi}$ will take all values in  $\mathcal{F}_k-\left\lbrace0\right\rbrace$ without any exclusion. Thus, the generalization of  \eqref{dcmp1} is
 \begin{equation}
 B(\phi,\phi')=\sum_{n=1}^{k-1}A_n(\phi,\phi').
 \end{equation}
 Let $\mathcal{L}_1=\left\lbrace1,2,\ldots,k-1\right\rbrace$. The set $\mathcal{L}_p$ for is constructed in the following manner: \\
 If $\mathcal{L}_p=\{ \phi_1,\phi_2,\ldots,\phi_{({k-1})^p}\} $, then  
 \begin{equation*}
 \mathcal{L}_{p+1}=\{ 1\phi_1,1\phi_2,\ldots,1\phi_{({k-1})^p},2\phi_1,2\phi_2,\ldots,2\phi_{({k-1})^p},\ldots,(k-1)\phi_1,(k-1)\phi_2,\ldots,(k-1)\phi_{({k-1})^p}\}.
 \end{equation*}
 where $1\phi_1$ is a configuration of states of $p+1$ vertical edges such that the state of the first edge is $1$ and the rest are in configuration $\phi_1$, and so on.
 For the case $k=3$, the sets $\mathcal{L}_p$ are the same as those defined in Section \ref{cylindricalice} with $1$ replaced by $\uparrow$ and $2$ replaced by $\downarrow$. If $\phi_i, \phi_j$ are the $i^{th}$ and $j^{th}$ elements of $\mathcal{L}_p$, then let
 \begin{equation}
 A_{n_p}(i,j)=A_n(\phi_i,\phi_j).
 \end{equation}
 Thus $A_{n_p}$ is a $(k-1)^p\times(k-1)^p$ matrix. Consider the subdivision of the matrix $A_{n_{p+1}}$ into $(k-1)^2$ block matrices each of size $(k-1)^p\times(k-1)^p$.\\ The $(l,m)^{th}$ element of the $(i,j)^{th}$ block matrix is 
 \begin{equation*}
 A_n(i\phi_l,j\phi_m)=A_{(-(k-n+k-j+i)\mod{k})}(\phi_l,\phi_m)=A_{[(n+j-i)\mod{k}]}(\phi_l,\phi_m)=A_{({(n+j-i)\mod{k})}_p}(l,m)
 \end{equation*}
 where $\phi_l,\phi_m \in \mathcal{L}_p$, and we have used  \eqref{arrow} and \eqref{icecondition}. Thus we get the following Toeplitz block structure
 \begin{equation}
 A_{n_{p+1}}=\begin{pmatrix}
 A_{n_p} & A_{{(n+1)}_p} & A_{{(n+2)}_p} & \ldots & \ldots & \ldots & \ldots & \ldots \\
 A_{{(n-1)}_p} & A_{n_p} & A_{{(n+1)}_p} & A_{{(n+2)}_p} & \ldots & \ldots & \ldots & \ldots \\
 A_{{(n-2)}_p} & A_{{(n-1)}_p} & A_{n_p} & A_{{(n+1)}_p} & A_{{(n+2)}_p} & \ldots & \ldots & \ldots \\
 \ldots & \ldots & \ldots & \ldots & \ldots & \ldots & \ldots & \ldots \\
 \ldots & \ldots & \ldots & \ldots & \ldots & \ldots & \ldots & \ldots \\
 \ldots & \ldots & \ldots & \ldots & \ldots & \ldots & \ldots & \ldots \\
 \ldots & \ldots & \ldots & \ldots & \ldots & \ldots & \ldots & \ldots \\
 \ldots & \ldots & \ldots & \ldots & \ldots & A_{{(n-2)}_p} & A_{{(n-1)}_p} & A_{n_p}
 \end{pmatrix},
 \end{equation}
 it being understood that the indices of the $A$ matrices are to be taken $\mod k$.
 
 \section{Graph-theoretical approach}
 
 Although graph-theoretical methods have been applied to the $k$-coloring problem (\cite{nagle,nagle1,avb}), the method we adopt for the $k$-coloring problem begins with the mapping between the $k$-coloring problem and the vertex model mentioned in  Section 2.4. Following  \cite{nagle}, $W_N$ is the number of ways of assigning states on the edges (of a very large two dimensional square lattice having $N$ lattice sites) consistent with the two rules \eqref{arrow} and \eqref{icecondition}: 
 \begin{equation}
 W_N = \sum_{[\xi]}\prod_{i<j}A(\xi_i,\xi_j).
 \end{equation}
 $A(\xi_i,\xi_j)$ is the compatibility matrix relating two adjacent sites. $\xi_i$ and $\xi_j$ are the configurations at two adjacent vertices. $\sum_{[\xi]}$ denotes the "sum over all different combinations of $\xi$ arrangements at each vertex $i$, and $\prod_{i<j}$ is the product over nearest neighbours with each pair taken once" \cite{nagle}. If $e_1$ and $e_2$ are the states of the edge (with respect to the two lattice sites connected by it) common to the two adjacent sites having configuration $\xi_i$ and $\xi_j$, then 
 \[A(\xi_i,\xi_j)=\begin{cases}
 1, & \text{if }e_1+e_2=0\mod{k} \\
 0, & \text{if }e_1+e_2\neq0\mod{k}
 \end{cases}
 \]
 
 The number of configurations ($M_k$, say) that a vertex can have is (here $|S|$ denotes the number of elements in set $S$)
 \begin{eqnarray*}
 M_k &=&\vert\left\lbrace(a_1,a_2,a_3,a_4):a_1+a_2+a_3+a_4=0\mod{k},a_1,a_2,a_3,a_4\in\mathcal{F}_k-\left\lbrace0\right\rbrace\right\rbrace\vert \nonumber \\
 &=&\vert\left\lbrace(a_1,a_2,a_3):a_1,a_2,a_3\in\mathcal{F}_k-\left\lbrace0\right\rbrace\right\rbrace\vert \nonumber \\ &-&  \vert\left\lbrace(a_1,a_2,a_3):a_1+a_2+a_3=0\mod{k},a_1,a_2,a_3\in\mathcal{F}_k-\left\lbrace0\right\rbrace\right\rbrace\vert
\nonumber \\
 &=& (k-1)^3-[(k-1)^2-\vert\left\lbrace(a_1,a_2):a_1+a_2=0\mod{k},a_1,a_2\in\mathcal{F}_k-\left\lbrace0\right\rbrace\right\rbrace\vert].
 \end{eqnarray*}
 So, 
 \begin{equation}
 M_k=(k-1)^3-(k-1)^2+(k-1).
 \end{equation}
 Thus the Pauling estimate for the $k$-coloring problem is $\frac{M_k}{(k-1)^2}$ and $\sum_{[\xi]}$ means sum over all possible $M_k^N$ configurations of $N$ lattice sites.
 We do the following transformation
 \begin{equation}\label{condition1}
 W_N=\sum_{[\xi]}\prod_{i<j}(C+A(\xi_i,\xi_j)-C)=(M_kC^2)^N\sum_{[\xi]}\prod_{i<j}\frac{1}{\sqrt{M_k}}(1+a(\xi_i,\xi_j))
 \end{equation}
 where 
 \begin{equation*}
 a(\xi_i,\xi_j)=\frac{1}{C}(A(\xi_i,\xi_j)-C).
 \end{equation*}
 Every term in the expansion of \eqref{condition1} can be assigned a graph (\cite{stillinger,nagle}). To cancel contributions from all open-ended graphs (\cite{stillinger,nagle}), the following condition must be satisfied  
 \begin{equation}\label{condition}
 \sum_{\xi_i(\text{or } \xi_j)}a(\xi_i,\xi_j)=0.
 \end{equation}
 If we fix $\xi_i$, then the number of possible $\xi_j$'s that will be compatible with $\xi_i$ is (where $b$ is the state of the edge connecting $\xi_i$ and $\xi_j$ with respect to the vertex $\xi_i$ and $a_1, a_2, a_3$ are the states of three of the four edges emanating from $\xi_j$ other than that connecting $\xi_i$ and $\xi_j$)
 \begin{eqnarray*}
 &~&\vert \left\lbrace (a_1,a_2,a_3):a_1+a_2+a_3=b\mod{k}, a_1,a_2,a_3 \in \mathcal{F}_k- \left\lbrace 0 \right\rbrace \right\rbrace\vert
 \nonumber \\
 &=&\vert\left\lbrace(a_1,a_2):a_1,a_2\in\mathcal{F}_k-\left\lbrace0\right\rbrace\right\rbrace\vert-\vert\left\lbrace a_1+a_2=b\mod{k}, a_1,a_2\in\mathcal{F}_k-\left\lbrace0\right\rbrace\right\rbrace\vert
 \nonumber \\
 &=& (k-1)^2 - (k-2).
 \end{eqnarray*}
 
 \begin{figure}
 \centering
 \includegraphics[width=0.5\textwidth]{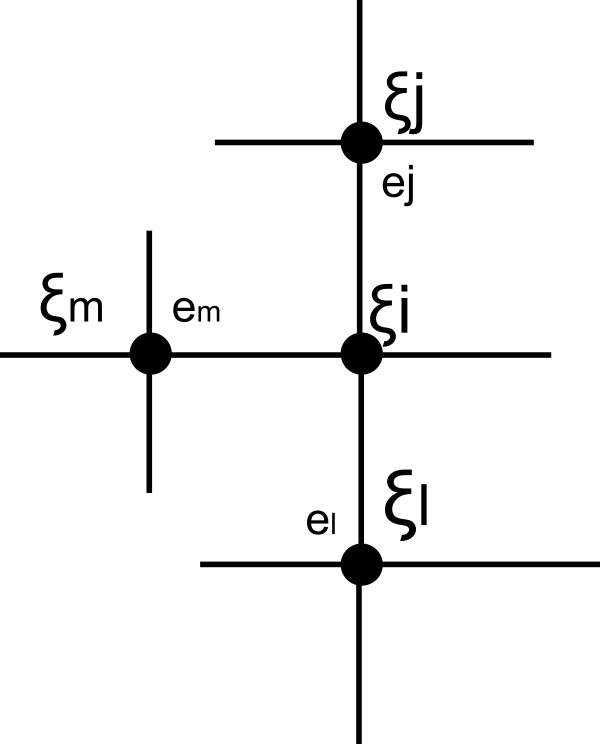}
 \caption{\label{fig:noneulerian}Three incident edges at the vertex having configuration $\xi_i$($e_j,e_l,e_m$ are the respective edge states with respect to vertices having configuration $\xi_j,\xi_l,\xi_m$ respectively).}
 \end{figure}
 
 Since the ones in $A(\xi_i,\xi_j)$ become $\frac{1-C}{C}$ in $a(\xi_i,\xi_j)$ and the zeros become $-1$,  \eqref{condition} gives 
 \begin{equation*}
 \left(\frac{1-C}{C}\right)[(k-1)^2-(k-2)]-(M_k-(k-1)^2+(k-2))=0.
 \end{equation*}
 Letting $k-1=x$, the equation written above simplifies to \begin{equation*}
 \frac{1}{C}=x=k-1
 \end{equation*}
 and \begin{equation*}
 \frac{1-C}{C}=x-1=k-2
 \end{equation*}
 and thus we get the following 
 \[a(\xi_i,\xi_j)=\begin{cases}
 (k-2), & \text{if }\xi_i \text{ and } \xi_j \text{ are compatible,} \\
 -1, & \text{if } \xi_i \text{ and } \xi_j \text{ are not compatible.}
 \end{cases}
 \]
 \subsection{Non-Eulerian cycles}
  We examine whether a generalization of a statement in \cite{nagle} that the contribution of non-Eulerian graphs, i.e., graphs having one or three incident edges at a vertex(\cite{nagle,berge}) have zero contribution, holds for a general $k$ or not. Graphs having one incident edge at a vertex have zero contribution because requirement \ref{condition} has already been fulfilled. Thus what remains to be examined(for a general $k$) is the value of the following expression for three incident edges 
  \begin{equation*}
 \sum_{\xi_i}a(\xi_i,\xi_j)a(\xi_i,\xi_l)a(\xi_i,\xi_m).
 \end{equation*}
 For the sake of notation, let (refer to Fig. \ref{fig:noneulerian}) $e_j=l_1$, $e_m=l_2$ and $e_l=l_3$. We have the following statements ($a_1, a_2, a_3$ are the states of the three edges connecting $\xi_i$ with $\xi_j, \xi_m, \xi_l$ with respect to the lattice site having configuration $\xi_i$, $a$ is the state of the remaining edge).
 Note that in the statements that follow, $a_i \in\mathcal{F}_k-\left\lbrace 0,(k-l_{i})\right\rbrace$, $a\in\mathcal{F}_k-\left\lbrace0\right\rbrace$.
 
 \begin{enumerate}
 \item  Number of configurations $\xi_i$ that are compatible with all three lattice sites is $1$ if $l_1+l_2+l_3\neq 0 \mod{k}$ and is $0$ otherwise.
 \item Number of configurations that are compatible with any two of the three lattice sites, and not with the third one(say the one out of which $l_3$ is protruding) is
 \begin{equation*}
 \vert\left\lbrace(a_3,a):a_3+a=l_1+l_2\mod{k}\right\rbrace\vert
 =(k-2)-\vert\left\lbrace a_3:a_3=l_1+l_2\mod{k}\right\rbrace\vert.
 \end{equation*}
 \item Number of configurations that are compatible with only one lattice site(say the one out of which $l_1$ is protruding) is
 \begin{eqnarray*}
 \vert\{ (a_2,a_3,a)&:&a_2+a_3+a=l_1\mod{k}\}\vert
 \nonumber \\
 &=&(k-2)^2-\vert\left\lbrace a_2+a_3=l_1\mod{k}\right\rbrace\vert
 \nonumber \\
 &=& (k-2)^2-[(k-2)-\vert\left\lbrace a_2:a_2=l_1\mod{k} \text{ or } a_2=l_1+l_3\mod{k}\right\rbrace\vert].
 \end{eqnarray*}
 \item Number of configurations that are compatible with none of the three lattice sites are 
 \begin{eqnarray*}
 &~&\vert \{ (a_1,a_2,a_3,a):a_1+a_2+a_3+a = 0\mod{k}\} \vert 
 =(k-2)^3-[(k-2)^2 \nonumber \\ &-&[(2(k-2)-\vert\left\lbrace a_1=l_2\mod{k}\right\rbrace\vert-\vert\left\lbrace a_1=l_3\mod {k}\text{ or }a_1=l_2+l_3\mod{k}\right\rbrace\vert]].
 \end{eqnarray*}
 \end{enumerate} 
 One can easily verify by the statements listed above that for the case $l_1+l_2+l_3=0\mod{k}$  
 \begin{equation*}
 \sum_{\xi_i}a(\xi_i,\xi_j)a(\xi_i,\xi_l)a(\xi_i,\xi_m)=(k-2)-2(k-2)^2-(k-2)^3+2. 
 \end{equation*}
 which is zero for $k=3$, which is Nagle's \cite{nagle} result, but for $k>3$, it is not zero. 
 
 \subsection{Elementary Eulerian Cycles}
 
 Our next step is to compute the following 
 \begin{equation*}
 \sum_{\xi_2, \xi_3, \ldots, \xi_{i-1}} a(\xi_1,\xi_2)a(\xi_2,\xi_3)\ldots a(\xi_{i-1},\xi_i) = a^{i-1}(\xi_1, \xi_i).
 \end{equation*}
 We use the symbol $a(\xi_i\leftrightarrow\xi_j)$ for $a(\xi_i, \xi_j)$, if $\xi_i$ and $\xi_j$ are compatible, otherwise we use $a(\xi_i \neq  \xi_j)$. The entries in the compatibility matrix $a$ clearly depend on the compatibility of the two adjacent sites, and the same is true for the higher powers of $a$. Thus we don't need to worry about what $\xi_i$ and $\xi_j$ are in $a(\xi_i \leftrightarrow \xi_j)$ or $a(\xi_i \neq \xi_j)$. We have the following recursion relation:
 \begin{eqnarray}
 a^i(\xi_i\leftrightarrow\xi_j) &=& \sum_{\xi_k}a^{i-1}(\xi_i,\xi_k)a(\xi_k,\xi_j) \nonumber \\
 &=&(k-1)a^{i-1}(\xi_1 \leftrightarrow \xi_2) a(\xi_1 \leftrightarrow \xi_2) + (k-2)^2 a^{i-1} (\xi_1 \leftrightarrow \xi_2) a(\xi_1 \neq \xi_2) \nonumber \\ &+& (k-2)^2 a^{i-1}(\xi_1 \neq \xi_2) a(\xi_1 \leftrightarrow \xi_2) + [(k-2)^3+(k-2)]a^{i-1}(\xi_1\neq\xi_2)a(\xi_1\neq\xi_2)\nonumber \\
 &=& (k-2)[a^{i-1}(\xi_1\leftrightarrow\xi_2)-a^{i-1}(\xi_1\neq\xi_2)].
 \end{eqnarray}
 Similarly, we have
 \begin{equation}
 a^i(\xi_1\neq\xi_2)=-(a^{i-1}(\xi_1\leftrightarrow\xi_2)-a^{i-1}(\xi_1\neq\xi_2))
 \end{equation}
 Fortunately, the recursion can be solved easily without resorting to any matrix method, and we get
 \begin{eqnarray*}
 a^i(\xi_1\leftrightarrow\xi_2) &=& (k-1)^{i-1}(k-2)
 \nonumber \\
 a^i(\xi_1\neq\xi_2) &=& -(k-1)^{i-1}
 \nonumber \\
 {\rm Tr}[a^i] &=& (k-1)^2a^i(\xi_1\leftrightarrow\xi_2)+[M_k-(k-1)^2]a^i(\xi_1\neq\xi_2)
\nonumber \\
 &=& (k-2)(k-1)^i.
 \end{eqnarray*}
 And so the contribution of elementary Eulerian cycle having $n$ vertices is 
\begin{equation*}
\frac{Tr[a^n]}{M_k^n}=\frac{(k-2)(k-1)^n}{M_k^n}.
\end{equation*}
At large values of $k$, their contribution goes as $\mathcal{O}(1/k^{2n-1})$.

\begin{figure}
\centering
\includegraphics[width=0.5\textwidth]{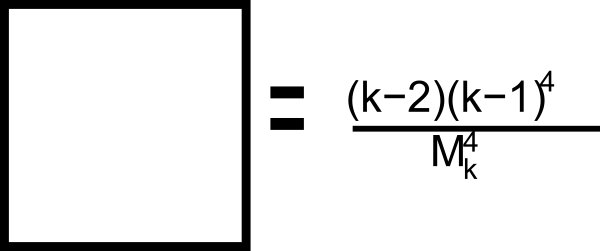}
\caption{Lowest order non-zero correction to Pauling type estimate for the $k$-coloring problem is due to elementary Eulerian cycle having $4$ vertices, which is a square.}
\end{figure}

For any cycle, we can say that the contribution of that cycle must go as $\mathcal{O}(1/k^n)$ (for some $n>0$), because if it were not so, the asymptotic relation, $W(k) \sim k$ for large $k$  \cite{nagle1} would not hold good. Thus, at large values of $k$, the mean field approximation (which is Pauling's estimate) dominates (refer also to Fig. 8): \begin{equation}
W(k) \approx \frac{M_k}{(k-1)^2}(1+\frac{(k-2)(k-1)^4}{M_k^4})\approx\frac{(k-1)^3-(k-1)^2+(k-1)}{(k-1)^2}.
\end{equation}
One can see that the above approximation for $k = 4, 5, 6, 7$ agrees well with the values given in \cite{nagle1}.
Calculating the general algebraic expressions for contributions due to complicated cycles is tricky and can be another avenue of investigation.

\section{Summary}

We have established the connection between $k$-coloring problem of the square lattice and a vertex model which reduces to the ice-type model for $k = 3$. We have, for instance, explicitly shown that 4-coloring problem is in correspondence with a 21-vertex model. We have defined a transfer matrix for the $k$-coloring problem and showed its Toeplitz block structure for the non-periodic boundary conditions. 

We have also calculated Lieb's square ice constant for cylindrical chain. Further, applying graph-theoretical methods to the vertex model corresponding to the $k$-coloring problem, we have shown  that the generalized Pauling estimate dominates for large values of $k$. 

\section*{Appendix}

\noindent
{\bf Analytical results for some finite models}
 
 We present some simple analytical results for models that remain finite in one of the dimensions using theorems stated in Section 2.2. 
 
\subsubsection*{A. ~$2\times p$ cylindrical lattice}

The number of ways, $Z_{2p}$ is

\begin{eqnarray}
Tr[B_p^2] &=& Tr[A_p^2+A_pA_p^T+A_p^TA_p+{A_p^T}^2] \nonumber \\ &=& 2(Tr[A_p^2]+Tr[A_pA_p^T]) = 2(2^p+3^p).
\end{eqnarray}
Thus, the maximum eigenvalue here is 
\begin{equation}
\lim_{p\rightarrow\infty}Z_{2p}^{1/{2p}}=\sqrt{3}\approx 1.732 \ldots  
 \end{equation}
 
 \subsubsection*{B. ~$3\times p$ cylindrical lattice}\label{3p}
 
The number of ways $Z_{3p}$ is
\begin{eqnarray}
Tr[B_p^3] &=& Tr[A_p^3+{A_p^T}^3+A_p^2A_p^T+A_pA_p^TA_p+A_p^TA_p^2+A_p{A_p^T}^2+A_p^TA_pA_p^T+{A_p^T}^2A_p] \nonumber \\ &=& 2Tr[A_p^3]+6Tr[A_p^2A_p^T].
\end{eqnarray}
Thus by using the two theorems, we get
 
\begin{equation}
Tr [B_p^3]=2^{p+1}+6\times 4^p.
\end{equation}
In this case, the maximum eigenvalue is 
 \begin{equation}
 \lim_{p\rightarrow\infty}Z_{3p}^{1/{3p}}=4^{1/3}\approx1.5874.
 \end{equation}
 
Evaluating $Z$ for $m \times p $ lattices for $m > 3$  is possible in principle and we get coupled recursion while evaluating terms like $Tr[A_p^2{A_p^T}^2]$ and in general, terms like $Tr[A_p^{m_1}{A_p^T}^{m_2}A_p^{m_3}\ldots]$.  The coefficients of these coupled recursion become difficult to predict  analytically for very large $m$. We present the evaluation for a $4\times p$ lattice as an illustration.

\subsubsection*{C.~$4 \times p$ cylindrical lattice
}
Before embarking on $Z_{4p}$, we evaluate $Tr[A_p^2{A_p^T}^2]$.
 \begin{eqnarray*}
 A_{p+1}^2{A_{p+1}^T}^2 &=& \begin{pmatrix}
 A_p^2 & A_pA_p^T+A_p^TA_p\\
 0 & A_p^2
 \end{pmatrix}
 \begin{pmatrix}
 {A_p^T}^2 & 0\\
 A_pA_p^T+A_p^TA_p & {A_p^T}^2 
 \end{pmatrix} \nonumber \\ \\ 
 &=& \begin{pmatrix}
 Y & A_p{A_p^T}^3+A_p^TA_p{A_p^T}^2\\
 A_p^3A_p^T+A_p^2A_p^TA_p & A_p^2{A_p^T}^2
 \end{pmatrix}
 \end{eqnarray*}
 where 
\begin{equation*} Y=A_p^2{A_p^T}^2+A_pA_p^TA_pA_p^T+A_p{A_p^T}^2A_p+A_p^TA_p^2A_p^T+A_p^TA_pA_p^TA_p. 
\end{equation*} 
Thus we get
\begin{equation}\label{coupled1}
 Tr[A_{p+1}^2{A_{p+1}^T}^2]= Tr[Y]+Tr[A_p^2{A_p^T}^2]\\
 =4Tr[A_p^2{A_p^T}^2]+2Tr[A_pA_p^TA_pA_p^T].
\end{equation}
We also have 
 \begin{eqnarray*}
 A_{p+1}A_{p+1}^TA_{p+1}A_{p+1}^T &=& \begin{pmatrix}
 A_p & A_p^T \\
 0 & A_p
 \end{pmatrix}
 \begin{pmatrix}
 A_p^T & 0\\
 A_p & A_p^T
 \end{pmatrix}
 \begin{pmatrix}
 A_p & A_p^T \\
 0 & A_p
 \end{pmatrix}
 \begin{pmatrix}
 A_p^T & 0\\
 A_p & A_p^T
 \end{pmatrix}
\nonumber \\ \\ 
 &=&\begin{pmatrix}
 A_pA_p^T+A_p^TA_p & {A_p^T}^2 \\
 A_p^2 & A_pA_p^T
 \end{pmatrix}
 \begin{pmatrix}
 A_pA_p^T+A_p^TA_p & {A_p^T}^2 \\
 A_p^2 & A_pA_p^T
 \end{pmatrix}
 \end{eqnarray*}
 Thus
 \begin{equation*}
Tr[A_{p+1}A_{p+1}^TA_{p+1}A_{p+1}^T] = Tr[(A_pA_p^T+A_p^TA_p)^2+{A_p^T}^2A_p^2]+Tr[A_p^2{A_p^T}^2+A_pA_p^TA_pA_p^T]
 \end{equation*}
 \begin{equation}\label{coupled2}
 =3Tr[A_pA_p^TA_pA_p^T]+4Tr[A_p^2{A_p^T}^2].
 \end{equation}
 Letting $x_p=Tr[A_p^2{A_p^T}^2]$ and $y_p=Tr[A_pA_p^TA_pA_p^T]$, Equations \ref{coupled1} and \ref{coupled2} reduce to the following coupled recursion
 \begin{equation*}
 x_0,y_0=1
 \end{equation*} 
 \begin{equation*}
 x_{p+1}=4x_p+2y_p
 \end{equation*} 
 \begin{equation}
 y_{p+1}=4x_p+3y_p.
 \end{equation}
 which is the same as 
 \begin{equation*}
 \begin{pmatrix}
 x_{p+1}\\
 y_{p+1}
 \end{pmatrix} =
 \begin{pmatrix}
 4 & 2\\
 4 & 3
 \end{pmatrix}
 \begin{pmatrix}
 x_p\\
 y_p
 \end{pmatrix}
 \end{equation*}
 and we get
 \begin{equation}
 \begin{pmatrix}
 x_p \\
 y_p
 \end{pmatrix}=
 \begin{pmatrix}
 4 & 2\\
 4 & 3
 \end{pmatrix}^p
 \begin{pmatrix}
 1\\
 1
 \end{pmatrix}
 \end{equation}
 The $p^{th}$ power was calculated using MATHEMATICA and (since $\frac{7+\sqrt{33}}{2}$ and $\frac{7-\sqrt{33}}{2}$ are the eigenvalues of the matrix) we get the following
 \begin{equation*}
 \begin{pmatrix}
 x_p\\ 
 y_p
 \end{pmatrix}
 =\begin{pmatrix}
 a_1(\frac{7+\sqrt{33}}{2})^p+a_2(\frac{7-\sqrt{33}}{2})^p\\
 b_1(\frac{7+\sqrt{33}}{2})^p+b_2(\frac{7-\sqrt{33}}{2})^p
 \end{pmatrix}
 \end{equation*}
 where $a_1,a_2,b_1,b_2$ are constants independent of $p$.
 The $Z_{4p}$ for the lattice is given by
 \begin{equation*}
 Tr[B_p^4]=Tr[(A_p+{A_p^T})^4]=2Tr[A_p^4]+8Tr[A_p^3A_p^T]+4Tr[A_p^2{A_p^T}^2]+2Tr[A_pA_p^TA_pA_p^T].
 \end{equation*}
 and thus \begin{equation}
 \lim_{p\rightarrow\infty}Z_{4p}^{1/{4p}}=(\frac{7+\sqrt{33}}{2})^{1/4} \approx 1.589.
 \end{equation}
 \subsubsection*{D.~Open Lattice}
 The number of ways $Z$ for open lattice is\begin{equation}
 Z=\sum_{\phi_{1}}\ldots \sum_{\phi_{m}}B(\phi_1,\phi_2)B(\phi_2,\phi_3)\ldots B(\phi_{m-1},\phi_m)=\sum_{\phi_1,\phi_m}B^{m-1}(\phi_1,\phi_m).
 \end{equation}
 Calculating $Z_{m\times p}$ for finite $m$ is still possible in principle by constructing recursive formula for sum of the elements of matrix $B_p^{m-1}$ similar to the way it is done for cylindrical lattice, but for large values of $m$, it becomes too involved.

\section*{Acknowledgements}
SRJ is very grateful to John William Turner for many illuminating discussions on mathematical physics during his visits to the Universit\`{e} Libre de Bruxelles, Belgium. One of those discussions focussed on iterative Block matrices and their possible connection with vertex models.  
This work was carried out when one of the authors (SKS) was visiting SRJ under the Summer Research Fellowship Programme organized by The Indian Academy of Sciences, The Indian National Science Academy and The National Academy of Sciences, India. SKS would like to thank the Academies for the support.

\end{document}